\documentclass[10pt,conference]{IEEEtran}
\usepackage{eurosym}
\usepackage{amssymb}
\usepackage{amsmath}
\usepackage{amsthm}
 \usepackage{paralist}
\usepackage{graphicx}
\usepackage[font=footnotesize, justification=centering, labelsep=period]{caption}
\usepackage [noadjust]{cite}
\newtheorem{def_}{Definition}
\begin{document}
\title{\textbf{\Large The Typed Graph Model}}
\author{
\IEEEauthorblockN{Fritz Laux}
\IEEEauthorblockA{Fakult\"{a}t Informatik\\
Reutlingen University\\
D-72762 Reutlingen, Germany\\
email: fritz.laux@fh-reutlingen.de}
}

\maketitle
\begin{abstract}
In recent years, the Graph Model has become increasingly popular, especially in the application domain of social networks. 
The model has been semantically augmented with properties and labels attached to the graph elements. 
It is difficult to ensure data quality for the properties and the data structure because the model does not need a schema. 
In this paper, we propose a schema bound Typed Graph Model with properties and labels. 
These enhancements improve not only data quality but also the quality of  graph analysis. 
The power of this model is provided by using hyper-nodes and hyper-edges, which allows to present a data structure on different abstraction levels.
We demonstrate by example the superiority of this model over the property graph data model of Hidders and other prevalent data models, namely the relational, object-oriented, and XML model. 
\end{abstract}

\IEEEpeerreviewmaketitle

\section{Introduction}
\label{sec:intro}

The popularity of the Graph Model (GM) stems primarily from its application to social networks. Commercial graph database products like Neo4J \cite{neo4j}, ArangoDB \cite{arango}, JanusGraph \cite{janus}, Amazon Neptune \cite{neptune}, and others have been successfully applied to many domains. There are applications to medicine, drug analysis, scientific literature analysis, power and telephone networks.

\begin{figure}[]
\centering
\includegraphics[width=0.25\textwidth]{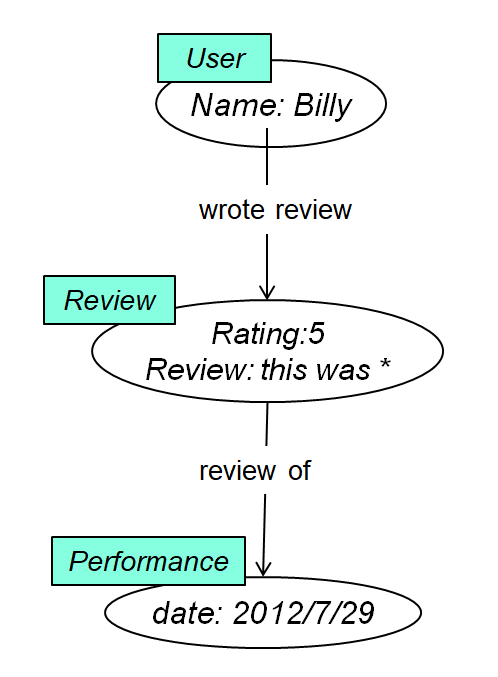}
\caption{Example graph taken partially from \cite{Robinson}, p. 42}
\label{fig:RobinsonsExample}
\end{figure}

The flexibility of the GM and its schema-less implementations are prone to data quality problems. 
Advocates of the GM like Robinson et al. of Neo4J recommend in their book \cite{Robinson} to use specification by example, which builds on example objects. 
But this reaches not far enough as the following example taken from Robinson's book shows. 
It is depicted in Figure \ref{fig:RobinsonsExample} and shows a \emph{User} named Billy with its 5-star  \emph{Review} on a \emph{Performance} dated 2012/7/29. 
From this example we cannot know if Billy is allowed to have multiple reviews (on the same performance).
For good data quality, a review should depend on the existence of a user and a performance. 
But this cannot be derived from one example. 
This means that we have to deal with class things (like a generic Person) and not only with real objects (like Billy) and specify if a relationship is mandatory or optional.

In order to express this information, it is necessary to abstract from a particular situation and specify integrity constraints. 
The use of a schema would help to ensure data integrity and would clarify the intended situation of the example.
Daniel et al. \cite{Daniel} also point out  the importance of a schema for data consistency and efficient implementation of a graph database.

Another weakness of the GM is that it has no notation to support different levels of detail and abstraction, which is apparently important for modeling large and complex data structures.

\subsection{Contribution}
To overcome these limitations we introduce in this paper a new typed graph model allowing hyper-nodes with complex structured properties (even sub-graphs) and hyper-edges connecting (recursively) one, two or more hyper-nodes. 
The graph schema provides data types, which allow type checking for instance elements. This ensures a formal data quality.
Our model has a higher semantic expressiveness and precision than the prevalent data models, namely the relational, object oriented, and XML data model. 
This will be demonstrated with typical modeling patterns.

\subsection{Structure of the Paper}
With the following overview of Related Work the context for our new typed graph model will be settled. 
Section \ref{sec:TypedGM} introduces and defines formally the Typed Graph Model (TGM) consisting of a typed schema and a hyper-graph instance connected to the schema. 
We present a compact and easy to read visualization of the model.
The definitions are illustrated by some examples. 
In the next Section \ref{sec:Comparison} our TGM is compared to the Graph Data Model (GDM) of J. Hidders \cite{Hidders}. 
Then, the semantic expressiveness of the TGM is demonstrated with typical data structures and
compared with the prevalent data models, namely the relational, object oriented, and XML data model.
The paper ends with a summary of our findings and gives an outlook on ideas for future work.

\section{Related Work}
Since the beginning of 1980 many papers on the GM have been published. DBLP \cite{dblp} alone retrieves 757 matches for the key words "graph data model". If we ignore the papers that present specific applications for the GM incl. XML or Hypertext applications a few dozen of relevant papers remain. In the following, we discuss only papers that present the GM and its extensions (e. g., the Property Graph Model (PGM)) with a formal foundation or papers that use a graph schema:

The notion of PGM was informally introduced by Rodriguez and Neubauer \cite{Rodriguez}.
Spyratos and Sugibuchi \cite{Spyratos} use property graphs with hyper-nodes and hyper-edges for their graph data model. 
The main difference to our approach is that no schema is used and properties have no predefined data type. 
Another approach with hyper-edges is presented by Bu et al. \cite{Bu} who treats a label like a node connecting a set of nodes, which he calls hyper-edge. The nodes itself can be of different types. In this case Bu calls the graph a unified hyper-graph. The unified hyper-graph model is then applied to problem of ranking music content and combining it with social media information. Compared to our TGM the unified hyper-graph of Bu is only defined for graph instances. It is not not clear if the nodes have any type checking and if the whole graph is ruled by a schema. 

Ghrab et al. \cite{Ghrab} present GRAB, a schemaless graph database based on the PGM. It supports integrity constraints but cannot ensure data quality because of missing data types for properties and labels. Neo4J \cite{Robinson} has similar foundations and features. It has optional support for integrity constraints and comes with a powerful and easy to use graph query language, called \textit{Cypher}. 

All these PGM originate as instance graphs and no special attention is given to the graph schema.
No attempt is made to specify the different types of edges and the multiplicity of connections (edges) between different node types. 
Nodes are not typed and labels are not a proper substitute. 

Amann and Scholl \cite{Amann} seem to be the first authors who connect a graph schema with its graph database instance. Nodes and edges do not have properties but both must conform to the schema. Their model is used for an algebra (hyperwalk algebra) for traversing the graph.

Marc Gyssens et al. \cite{Gyssens} and Jan Hidders \cite{Hidders} use a labeled GM to represent a database schema where each property of an object is modeled as a node in the graph. 
Labels are used to name node classes and edges.
The models become confusing because a node represents either an object, a property or a data type. 
Still, it is not possible to restrict the cardinality of schema edges (relationships).
Hidders' model is explained in more detail and compared to our TGM in Section \ref{sec:Comparison}.

Similar to Amann and Scholl the paper of Pab\'on et al. \cite{Pabon} uses a graph schema to query the graph database. 
They distinguish different node types, which they call "sort". 
The supported types are: \textit{object class nodes} (complex objects), \textit{composite-value class nodes} (for aggregate values), and \textit{basic-value class nodes} (primitive data types). 
This model seems to be equivalent to (complex) nodes with properties governed by a schema. 
A mechanism to abstract and group sub-graphs would help to make the model easier to communicate.

Pokorn\'y \cite{Pokorny} uses a binary ER-Model as graph conceptual schema. 
For the graphical rendering he uses a compact entity representation for the nodes with attribute names inside the entity box. 
This solves the problem using the same node symbol for entities and attributes (properties) as it is the case with Gyssens \cite{Gyssens} and Hidders \cite{Hidders} models. 
The edge cardinality is represented in a form of crow-foot notation. 

In order to make the GM usable for real life scenarios with hundreds of schema elements, it is necessary to group or combine graph elements to higher abstracted objects. This would make the model easier to handle.

The need for grouping graph elements is addressed by Junghanns et al. \cite{Junghanns}. 
Their model allows to form logical sub-graphs (graph collections) with heterogeneous nodes and edges. 
With this it is possible to aggregate sub-graphs, e. g., user communities. 
The authors use UML-like graphical rendering of nodes to make the model better readable but their model fails to specify the cardinality of schema edges.

A step toward to complex composite nodes as an alternative approach to aggregation presents Levene \cite{Levene} by allowing the graph vertices to be recursively defined as a finite set of graphs. These hyper-nodes do not form a well-founded set as a node may contain itself, which violates the foundation axiom for the Zermelo-Fraenkel set theory. 

 A relatively new formal definition including integrity constraints was given by Angles \cite{Angles}. However, his model does not allow structured objects and grouping or aggregation. In the following section, we simplify his definitions and use it as basis for our TGM. 
 
\subsection{Comparison with Ontology Languages}
Ontology languages like RDFS \cite{RDFS} and OWL \cite{OWL} are designed to specify ontologies and have their strength in allowing reasoning over instances of it. They are often used to semantically describe Linked Open Data (LOD) and the statement triples are usually visualized as graph structures.
RDFS and OWL provide a general type system that could be used to form user defined types. This would allow to use it as basis for a graph schema language. But if we look at the W3C OWL 2 Structural Specification \cite{OWLstructSpec} it seems difficult to define user specific classes and W3C itself uses UML class diagrams to illustrate OWL structures. 

The specification of data structures is not their core intention. In RDFS for instance it is not possible to define the cardinality of relationships. Likewise, OWL Lite has strong limitations on allowing only 0 or 1 as multiplicity of properties. Simple unique requirements and relations like one-to-one, one-to-many and many-to-one are cumbersome to define even in OWL Full.
Complex data structures need a modeling language that allows to define different levels of abstraction, which is not the strength of these ontology languages. Most examples of RDFS or OWL do not care about the multiplicity of relationships (cardinalities may be guessed via property names) and grouping of attributes seems to be on the same level as objects or subjects. 

\section{The Typed Graph Model}
\label{sec:TypedGM}

Our TGM informally constitutes a directed property hyper-graph that conforms to a schema.
In the following definitions our notation uses small letters for elements (nodes, edges, data types, etc.) and capital letters for sets of elements. Sets of sets are printed as bold capital letters. A typical example
would be $n \in N \in \mathbf{N} \subseteq \wp(N)$, where $\wp(N)$ is the power-set of $N$.

\subsection{Graph Schema}
\label{ssec:Schema}
 
Let $T$ denote a set of simple or structured (complex) data types. 
A data type $t := (l,d) \in T$ has a name $l$ and a definition $d$. 
Examples of simple (predefined) types are $(int, \mathbb{Z})$, $(char, ASCII)$, etc.
It is also possible to define complex data types like an order line $(OrderLine, (posNo, partNo,  partDescription, quantity))$. 
The components need to be defined in $T$ as well, e. g., $(posNo, int > 0)$. 
Recursion is allowed as long as the defined structure has a finite number of components.  

\begin{def_}[Typed Graph Schema]
A typed graph schema is a tuple $TGS = (N_S, E_S, \rho, T, \tau, C)$ where:
\begin{itemize}
\item
$N_S$ is the set of named (labeled) objects (nodes) $n$ with data type $t := (l,d) \in T$, where $l$ is the label and $d$ the data type definition.
\item
$E_S$ is the set of named (labeled) edges $e$ with a structured property $p := (l,d)\in T$, where $l$ is the label and $d$ the data type definition.
\item
$\rho$ is a function that associates each edge $e$ to a pair of object sets $(O,A)$, i. e., $\rho(e) := (O_e,A_e)$ with $O_e,A_e \in \wp(N_S)$. $O_e$ is called the \emph{tail} and $A_e$ is called the \emph{head} of an edge $e$.
\item
$\tau$ is a function that assigns for each node $n$ of an edge $e$ a pair of positive integers $(i_n,k_n)$, i. e., $\tau_e(n) := (i_n, k_n)$ with $i_n \in \mathbb{N}_0$ and $k_n \in \mathbb{N}$. 
The function $\tau$ defines the min-max multiplicity of an edge connection. 
If the min-value $i_n$ is $0$ then the connection is optional.
\item
$C$ is a set of integrity constraints, which the graph database must obey.
\end{itemize}
\end{def_}

The notation for defining data types T, which are used for node types $N_S$ and edge types $E_S$, can be freely chosen. This makes the expressiveness of the TGS at least as strong as the models to which it is compared in Section \ref{sec:Comparison}. 

\subsection{Typed Graph Model}
\label{ssec:TGM}

\begin{def_}[Typed Graph Model]
A typed graph Model is a tuple $TGM = (N, E, TGS, \phi)$ where:
\begin{itemize}
\item
$N$ is the set of named (labeled) nodes $n$ with data types from $N_S$ of schema TGS.
\item
$E$ is the set of named (labeled) edges $e$ with properties of types from $E_S$ of schema TGS.
\item
$TGS$ is a typed graph schema as defined in Subsection \ref{ssec:Schema}.
\item
$\phi$ is a homomorphism that maps each node $n$ and edge $e$ of $TGM$ to the corresponding type element of $TGS$, formally:
\begin{eqnarray*}
 \phi: & TGM &\rightarrow TGS \\
 & n &\mapsto  \phi(n) := n_S (\in N_S)\\
 & e &\mapsto  \phi(e) := e_S (\in E_S)
\end{eqnarray*}
\end{itemize}
\end{def_}
The fact that $\phi$ maps each element (node or edge) to exactly one data type implies that each element of the graph model has a well defined data type. 
The homomorphism is structure preserving. This means that the cardinality of the edge types are enforced, too. Data type and constraint checking is applied for all nodes and edges before any insert, update, or delete action can be committed. If no single type can be defined, union type or \emph{anyType} (sometimes called \emph{variant}) may be applied. Usually this is an indication for a weak data model and it should be clear that this could affect data quality and processing. 

As graphical representation for the TGS we adopt the UML-notation for nodes and include the properties as attributes including their data type. Labels are written in the top compartment of the UML-class.
Edges of the TGS are represented by UML associations.
For the label and properties of an edge we use the
UML-association class, which has the same rendering as an ordinary class but its existence depends on an association (edge), which is indicated by a dotted line from the association class to the edge. This not only allows to label an edge but to define user defined edge types.
The correspondence between the UML notation and the TGS definition is the following:\\

\begin{table} [h]
\caption{TGS correspondence with UML notation}\label{TGSvsUML}
\centering
\begin{tabular}{l | l}
TGS & UML \\
\hline
$n \in N_S$ & class \\
$e \in E_S$ & association \\
$t=(l,d) \in T$ & $l$ = name of $n$ resp. $e$; $d$ = type of $n$ resp. $e$ \\
$\rho(e)$ & all ends of $e$ \\
$\tau_e(n)$ & (min,max)-cardinality of e at n \\
$C$ & constraints in [ ] or \{ \} \\
\end{tabular}
\end{table}

The use of hyper-nodes $n \in N_S$ and hyper-edges $e \in E_S$ instead of simple nodes resp. edges allow to group nodes and edges to higher abstracted complex model aggregates. This is particularly useful to keep large models clearly represented and manageable.
Large graph models may then be grouped into sub-graphs like in Junghanns et al.\cite{Junghanns}. 
Each sub-graph can be rendered as a hyper-node. 
If the division is disjoint these hyper-nodes are connected via hyper-edges forming a higher abstraction level schema (see Figure \ref{fig:Enterprise} (b)).

\subsection{Examples}
\label{ssec:Examples}
\begin{figure}[]
\centering
\includegraphics[width=0.48\textwidth]{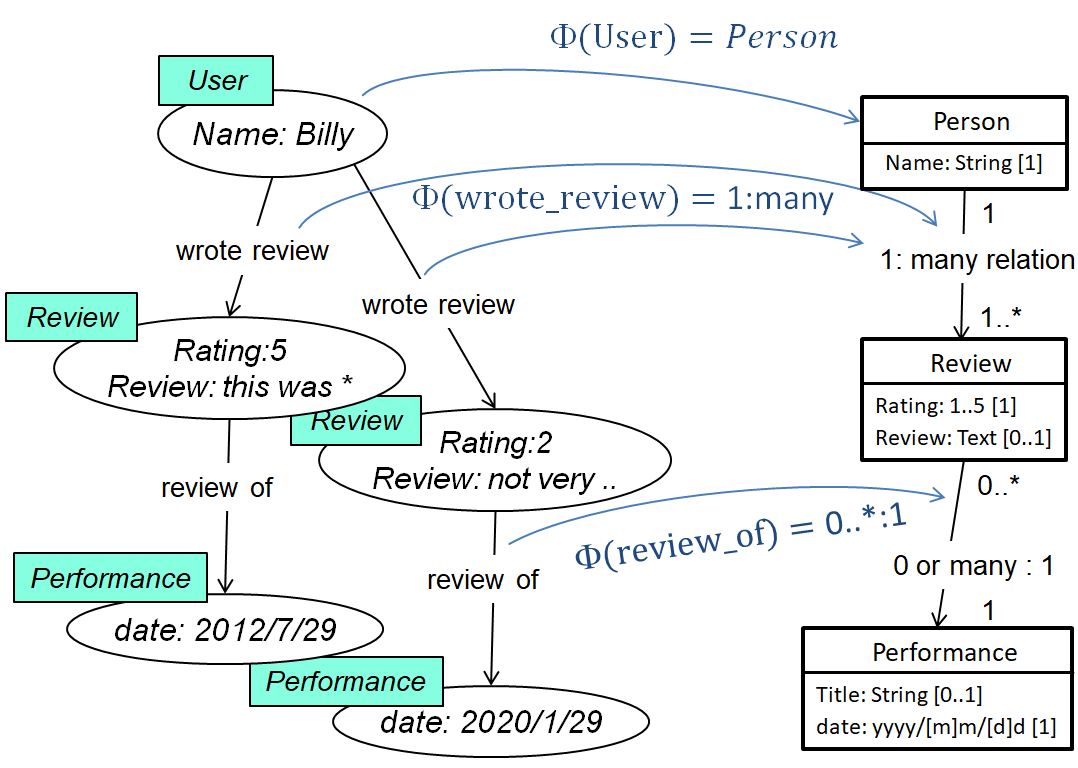}
\caption{Example graph with schema in UML notation}
\label{fig:SchemaRobinsons}
\end{figure}

Lets recall the example graph from Figure \ref{fig:RobinsonsExample} and model its corresponding schema. 
We want to make clear that a user may write as many reviews as he likes, but only one for a particular performance. 
A rating needs to refer exactly to one performance and one user. This is reflected in Figure \ref{fig:SchemaRobinsons} by the "1:many" and "0 or many:1" relationships.
We use the UML-notation for the schema and keep the notation from Figure \ref{fig:RobinsonsExample} for the instance graph for clarity.

The homomorphic mapping $\phi$ guaranties that the instance graph obeys the schema, i. e. type, cardinality, and constraint checking.
Now, it is clear from the schema that a user must have at least one review. The review is existence dependent on the user and a performance. The "wrote review" edge is a 1:many relation and "review of" is an optional many:1 relation. This has the consequence that a review needs a person and a performance. But, a performance may exist without any review.

\begin{figure*}[]
\centering
\includegraphics[width=0.7\textwidth]{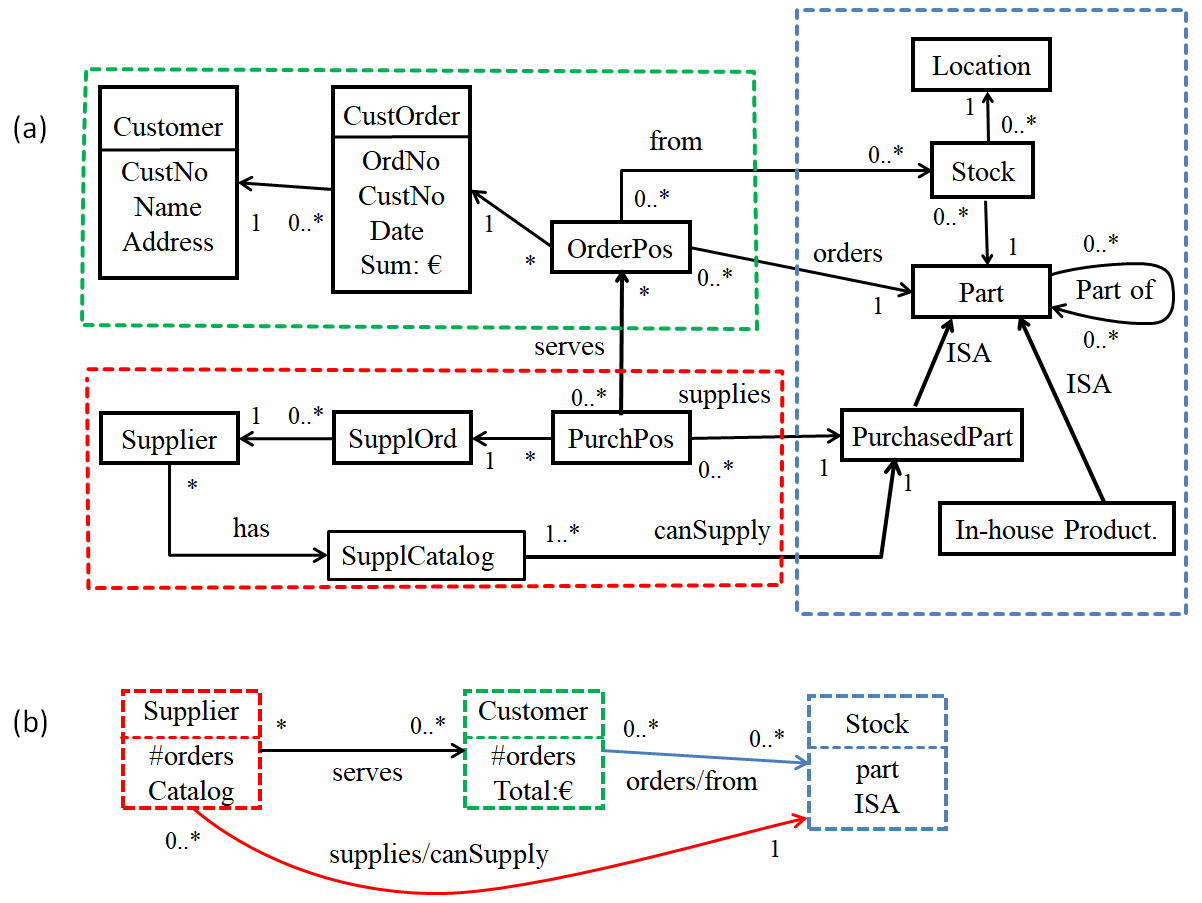}
\caption{Example TGM of a commercial enterprise showing two levels of detail}
\label{fig:Enterprise}
\end{figure*}

In the next example we present a commercial enterprise that sells products and parts to customers. 
The enterprise assembles products from parts and if the stock level is not sufficient it purchases parts from different suppliers.
Figure \ref{fig:Enterprise} models this situation using UML rendering. It demonstrates the abstraction power of the TGM showing two schema abstraction levels. 
The upper part (a) shows the TGM on a detailed level. 
The properties are suppressed in the diagram for simplicity except for \emph{Customer} and \emph{CustOrder}. The schema is grouped into 3 disjoint sub-graphs depicted with dashed shapes. 

In the lower part (b) these sub-graphs are shown as hyper-nodes of the graph schema. 
This allows a simplified and more abstracted view of the model. 
Also, some aggregate properties (e. g. \#orders) are shown to illustrate the modeling capabilities. 
The hyper-edges connecting these abstracted nodes must use the most general multiplicity of the multiple edges it combines. 
In the example the edge \emph{orders/from} combines two edges, i. e., \emph{orders} with 0..1 - 1 multiplicity and \emph{from} with 0..* - 0..* multiplicity, which leads to the most general multiplicity.

\section{Comparison with other Data Models}
\label{sec:Comparison}

In the following, we compare our TGM to other models with respect to structural differences and schema support. We point out modeling restrictions of these models and show how such situations are modeled with TGM. Query and manipulation languages are beyond the scope of this paper.

\subsection{Comparison with GDM of Jan Hidders}
\label{ssec:Hidders}

Jan Hidders' \cite{Hidders} model added labels and properties together with their data types to nodes and edges  (relationships).
Property names are modeled as edges in the schema.
This allows to model labeled relationships 
with complex properties. 
Structured and base data types share the same graphical representation, which makes it difficult to distinguish both. 
The ISA-relationship is rendered as a double line arrow. 
Hidders' model does not allow to restrict the cardinality of relationships. This restriction limits its modeling power compared to the TGM, which provides a min-max notation for the cardinality.

\begin{figure}[]
\centering
\includegraphics[width=0.45\textwidth]{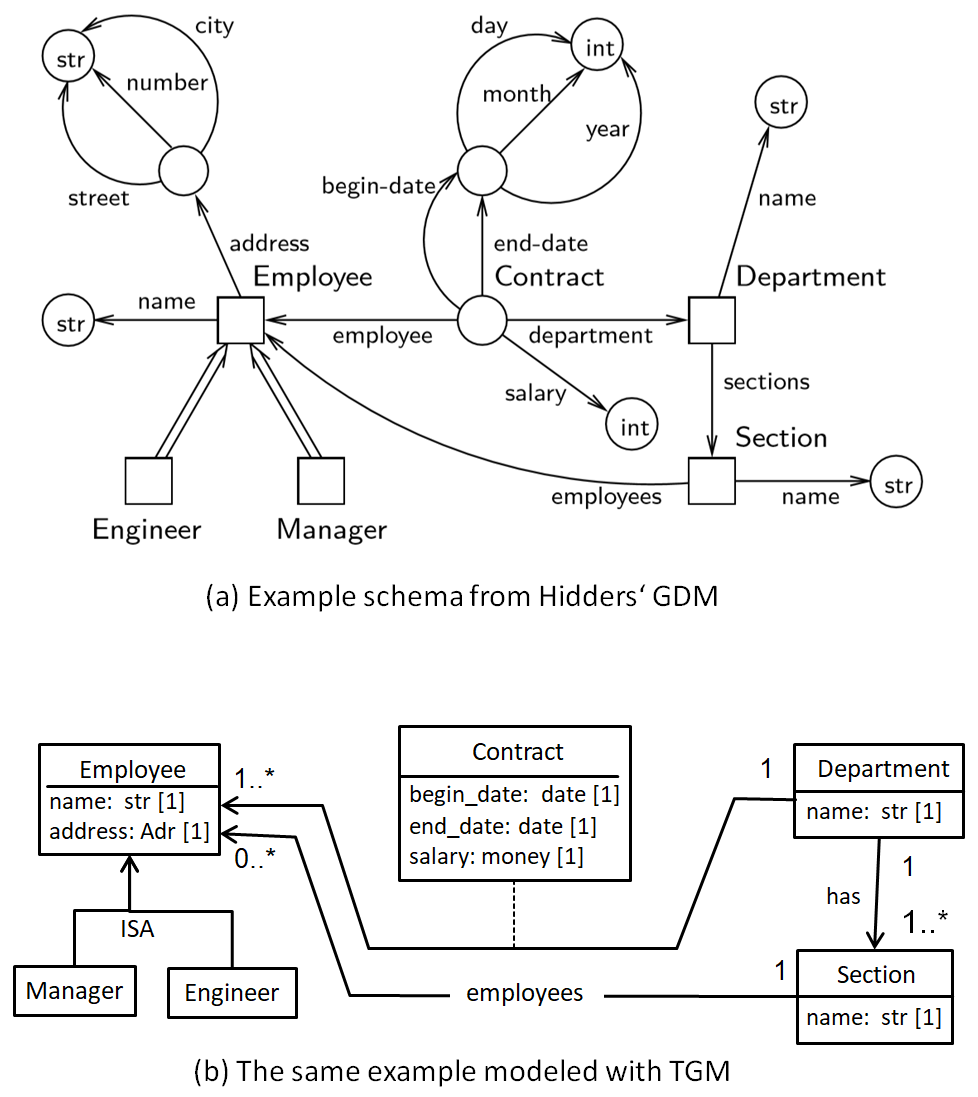}
\caption{Comparison by example with Hidders' GDM}
\label{fig:CompareHidders}
\end{figure}

The example in Figure \ref{fig:CompareHidders} is from the publication of Hidders \cite{Hidders}. 
The schema shows \emph{Employee} and \emph{Department} classes linked by a \emph{Contract}. The relationship \emph{Contract} is existence dependent on the connected nodes. The properties of \emph{Contract} are salary of type \emph{int}, begin-date and end-date of structure-type $date = (day, month, year)$.
In Hidders' model these dates are modeled on the element level using data type \emph{int}.
Hidders' schema elements, i. e., nodes (objects), edges (properties) and data types appear on the same visual level, which makes it difficult to read and obscures semantics. The modeling power of complex data types provide a clear advantage for the TGM.

\subsection{Comparison with the Relational Model (RM)}
\label{ssec:RM}

\begin{figure}[]
\centering
\includegraphics[width=0.3\textwidth]{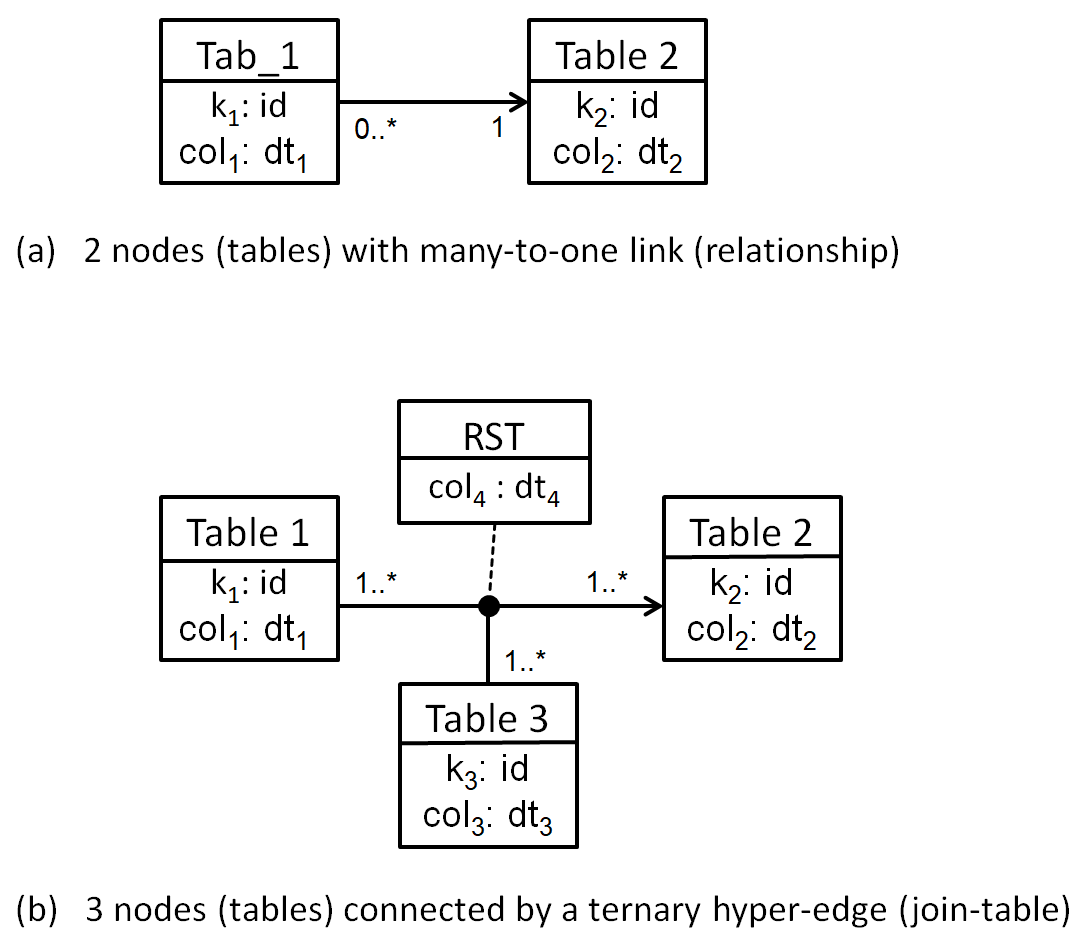}
\caption{Modeling a many-to-one relationship (FK) and a ternary join-table with TGM}
\label{fig:CompareRM}
\end{figure}

There is a 1:1 correspondence between attributes and properties and any relation can be modeled as a node  with properties. 
The min-max notation for relationship multiplicity can model any link cardinality.
The TGM can therefore easily represent tabular structures, foreign key constraints (many-to-one relationships), and join-tables as the building blocks of the RM. 
Beyond this, the TGM is able to directly model many-to-many relationships of any min-max multiplicity. This makes the TGM strictly stronger than the relational model.
Another difference to the RM is that foreign keys (FK) are not necessary because their function is taken over by an edge linking the FK-node (Table 1 without FK) with the referenced node (Table 2).
This can be seen in Figure \ref{fig:CompareRM} (a).

A join-table in the RM is existence dependent on the tables it refers to by FKs. 
The FKs forming the primary key (PK) of the join table are not necessary in the TGM because of the same reason as mentioned above.

In Figure \ref{fig:CompareRM} (b) the join-table RST maps directly to an hyper-edge labeled RST with property $col_3$ and without FKs. 
To make the ternary relationship example less abstract the RST could be an offer of products from Table 1 from a supplier of Table 3 to the client of Table 2. 
With this in mind it is clear that an offer depends on the product, the supplier, and the client.

The TGM can also represent non-normalized tables because the model supports complex structured data types having multivalued or array data. 
It is only necessary to define the necessary data types in the set of available data types $T$.

\subsection{Comparison with XML Schema}
\label{ssec:XML}

XML documents represent hierarchical hypertext documents. 
The document structure is defined by an XML schema. The hierarchy of XML-documents is directly supported by the TGM using directed edges. XLink provides references (arcs) between elements of internal or external XML-documents. Extended XLinks can connect to more than one element, but the references are always instance based, i. e. the target elements must be listed by URI. 
The TGM is more abstract and expressive allowing the definition of non-hierarchical references on the schema level.

As example serves a bookstore offering an unlimited number of books.
A simple XML-schema for the bookstore is given by w3schools.com.
The schema defines books with elements like "title", "author", etc. and its corresponding data types. 
Some data types are not as precise as they could, e. g. the data type xs:double for the price element. We will replace cs:double in our TGM by the money-type \emph{euro} to be more precise.
Some elements have attributes attached like the language ("lang") of a book title. 
The attribute minOccurs="1" of xs:sequence requires the bookstore to have a least one book.

\begin{footnotesize}
\begin{verbatim}
<?xml version="1.0" encoding="utf-8"?>
<xs:schema ... >
  <xs:element name="bookstore" >
    <xs:complexType >
      <xs:sequence minOccurs="1" 
          maxOccurs ="unbounded" >
        <xs:element name="book" >
          <xs:complexType>
            <xs:sequence>
              <xs:element name="title" >
              <xs:complexType>
                <xs:simpleContent>
                  <xs:extension base="xs:string">
                    <xs:attribute name="lang"
                        type="xs:string" />
                  </xs:extension>
                </xs:simpleContent>
              </xs:complexType>
                </xs:element>
              <xs:element name="author" 
                  type ="xs:string"/>
              <xs:element name="year" 
                  type ="xs:integer"/>
              <xs:element name="price" 
                  type ="xs:double"/>
            </xs:sequence>
            <xs:attribute name="category"
                type="xs:string"/>
          </xs:complexType>
        </xs:element>
      </xs:sequence>
    </xs:complexType>
  </xs:element>
</xs:schema>
\end{verbatim}
\end{footnotesize}

If we model the XML-elements as nodes in TGM then XML-attributes and the element values should be represented as properties. 
The name of an XML-element is mapped to a node label. The order of the XML-elements cannot be represented with this approach and XML-element values can be distinguished from XML-attributes by convention only. 

An alternative TGM model represents the complete book structure as one node. 
In this case the XML-elements and their attributes are modeled as structured properties of the book. 
The order of the elements and their associated attributes can be preserved. In fact, if XML Schema is used for specifying the data types $N_S$ and $E_S$ (see Subsection \ref{ssec:Schema}) all the flexibility and semantics provided be XML Schema can be represented with the TGS. This argument shows that the TGM is at least as powerful as the XML model.

\begin{figure}[]
\centering
\includegraphics[width=0.48\textwidth]{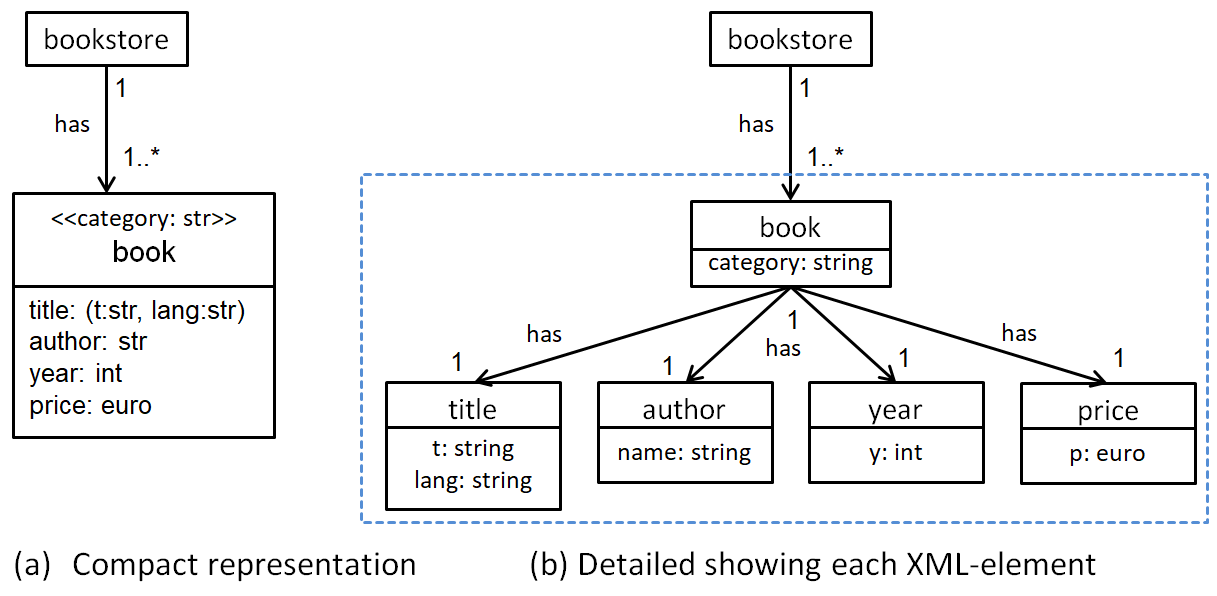}
\caption{Comparison by example with the XML Schema}
\label{fig:CompareXML}
\end{figure}

The example bookstore is depicted in Figure \ref{fig:CompareXML} where the left part (a) shows the compact version with the whole book modeled as one node and the right part (b) shows the version where each XML-element is modeled as node. 
We see from this example another possibility to use sub-graphs for higher abstracted models. 

\subsection{Comparison with the Object-Oriented Model}
\label{ssec:OOM}

Because we already use the UML for rendering the TGM, it is easy to see that classes correspond one-to-one with typed hyper-nodes. 
Any methods are simply ignored as we only deal with the network structure of OOM. 
Any complex internal class structure can be directly modeled by appropriate data types $t \in T$. 
The type set $T$ is defined beforehand but can contain any user defined structures. 
In contrast to the OOM the TGM allows different levels of abstraction in the modeling depending whether a structure is modeled by a detailed graph with simple types or a more compact graph using complex data types. This shows the same semantic expressiveness for structures, but a higher flexibility of the TGM. Considering the operations on data the OOM has the advantage to specify the allowed operations by methods.

The UML provides a rich set of association types, which need to be mapped to the label of the edges. 
Our TGM provides types not only for nodes but also for edges (called associations in UML). 
With this information it is possible to model different association types like aggregation, generalization, etc. 
Even user defined associations are possible, e. g., an aggregate could be further qualified as un-detachable or detachable composition or a loose containment.  
The arrow of the edge only indicates the reading direction of the association but does not limit the navigation of the TGM.

\begin{figure}[]
\centering
\includegraphics[width=0.48\textwidth]{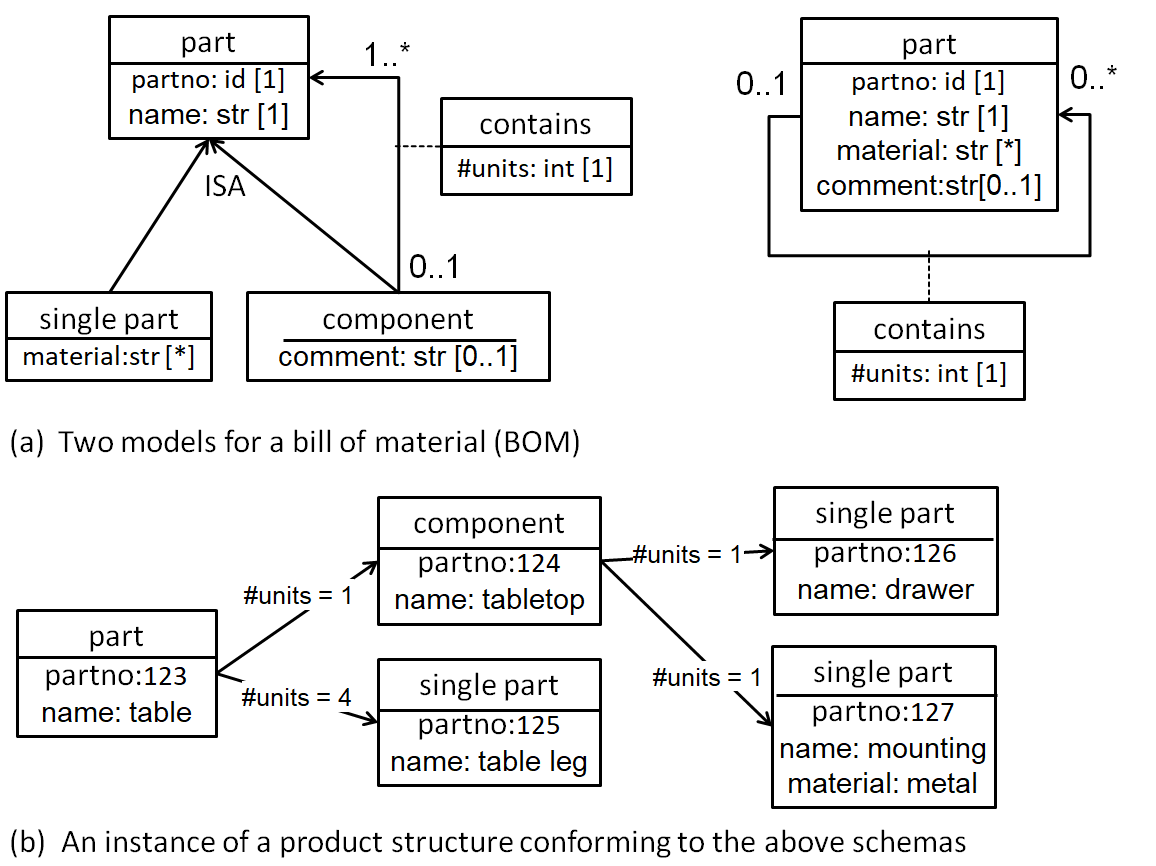}
\caption{Comparison by example with the OOM}
\label{fig:CompareOOM}
\end{figure}

It is also possible to model recursive structures as the examples from Figure \ref{fig:CompareOOM} illustrates. The bill of material (BOM) is an important example for a recursive structure used in production planning and control. It defines recursively a (compound) part with its components until a single part is reached. As example a table is given in Figure \ref{fig:CompareOOM} (b) consisting of 4 table legs and a tabletop consisting of a drawer and a mounting.

If the edge of \emph{contains} in Figure \ref{fig:CompareOOM} (a) is followed against the arrow direction it is possible to find the component where an individual part is built-in.
A complete \textit{where-used list} for a generic (not an individual) part may be obtained with a small schema modification. 
The from-end of the \emph{contains}-edge needs to change its multiplicity from 0..1 to 0..*. With this modification all components can be identified where a generic part is used. 
  
\section{Conclusion and Future Work}
This paper presents a structure definition of the TGM and an UML-like notation to visualize a graph database and its graph schema. 
Due to the TGS with predefined and user-defined data types, the TGM improves the formal data quality compared to other graph models.
We have demonstrated the superior modeling power in comparison to other graph data models and prevalent data models, namely relational, object oriented and XML model.
The model supports built in and user defined complex data types, which allow different abstraction levels. 
Another possibility for abstraction is to compress a sub-graph into a hyper-node reducing the visible complexity.
   
Because of its semantic modeling power the TGM could serve as a model that supports data integration from various data sources with different data models. 
The main challenge for an automated data integration are incompatible data sources where the TGM could help to solve quality issues and resolve inconsistent data.
Details still need to be investigated.
The development of a manipulation and query language for the TGM is future work. 
The idea is to combine elements of other graph languages with the dot-notation known from object-oriented languages.

\end{document}